\documentclass[aps,prd,reprint,preprintnumbers,groupedaddress,nofootinbib]{revtex4-1}

\usepackage{amsmath}

\newcommand{\be}{\begin{equation}}
\newcommand{\ee}{\end{equation}}

\DeclareRobustCommand{\Sec}[1]{Sec.~\ref{#1}}

\DeclareRobustCommand{\Tab}[1]{Table~\ref{#1}}

\DeclareRobustCommand{\Eq}[1]{(\ref{#1})}

\DeclareRobustCommand{\Ref}[1]{Ref.~\cite{#1}}

\newcommand{\Tau}{\mathcal{T}}

\begin{document}

\preprint{MIT-CTP 4685}

\title{Separated at Birth:  Jet Maximization, Axis Minimization, and Stable Cone Finding}

\author{Jesse Thaler}
\email[]{jthaler@mit.edu}
\affiliation{Center for Theoretical Physics, Massachusetts Institute of Technology, Cambridge, Massachusetts 02139, USA}

\begin{abstract}
Jet finding is a type of optimization problem, where hadrons from a high-energy collision event are grouped into jets based on a clustering criterion.  As three interesting examples, one can form a jet cluster that (1) optimizes the overall jet four-vector, (2) optimizes the jet axis, or (3) aligns the jet axis with the jet four-vector.  In this paper, we show that these three approaches to jet finding, despite being philosophically quite different, can be regarded as descendants of a mother optimization problem.  For the special case of finding a single cone jet of fixed opening angle, the three approaches are genuinely identical when defined appropriately, and the result is a stable cone jet with the largest value of a quantity $J$.  This relationship is only approximate for cone jets in the rapidity-azimuth plane, as used at the Large Hadron Collider, though the differences are mild for small radius jets.
\end{abstract}


\maketitle

\section{Introduction}

Jet algorithms are essential tools for connecting long-distance measurements made on hadrons to short-distance interpretations based on perturbative quarks and gluons.  Since there is a fundamental mismatch between color-singlet hadrons and color-carrying partons, there is no way to define an ideal jet finding procedure.  For this reason, a variety of jet algorithms have been introduced with different underlying philosophies and different practical advantages \cite{Ellis:2007ib,Salam:2009jx}.

In this paper, we expose a surprising connection between three seemingly unrelated approaches to jet finding:  jet function maximization \cite{Georgi:2014zwa,Ge:2014ova,Bai:2014qca}, 1-jettiness minimization \cite{Stewart:2010tn,Thaler:2010tr,Thaler:2011gf,Jouttenus:2013hs}, and stable cone finding \cite{Blazey:2000qt,Ellis:2001aa,Salam:2007xv}, all reviewed in \Sec{sec:review}.  Philosophically, these algorithms are quite different, so one might think that they would yield rather different jets.  Instead, all three algorithms yield (approximately) conical jets of (approximately) fixed radius $R$, with a high degree of correlation between the methods.  As we will show, this correlation is not an accident, since jet function maximization and 1-jettiness minimization can be viewed as descendants of a mother optimization problem, whose solution is a stable cone jet.

This relationship is most transparent in electron-positron collisions, where jets are typically defined in terms of particle energies and angles.  Remarkably, with appropriate definitions, all three algorithms can be made to yield \emph{identical} cone jets.  We will prove this in \Sec{sec:meta} by showing how these three methods can be derived from optimizing a common meta function.  This optimization is equivalent to finding all stable cones and then choosing the one with the largest value of $J$ (defined below).  In \Sec{sec:twoparticle}, we explain this relationship in more detail by performing a two particle case study.

Turning to the Large Hadron Collider (LHC), the intrinsic differences between the three algorithms become apparent.  In proton-proton collisions, jets are typically defined in terms of particle transverse momenta and rapidity-azimuth distances.  As discussed in \Sec{sec:LHC}, the transverse momentum of a jet is not exactly the same as the summed transverse momenta of its constituents.  For this reason, there is now a difference between optimizing a four-vector (as in jet functions), optimizing a light-like axis (as in 1-jettiness), and aligning the jet axis with the jet momentum (as in stable cones).  Despite these differences, though, we show that the algorithms still give similar results when the jet radius $R$ is small.

\section{Review of Jet Algorithms}
\label{sec:review}

To begin, we briefly review these three approaches to jet finding, all of which are infrared and collinear (IRC) safe.  For simplicity, in this paper we only consider finding the single hardest jet in an event, though all of these methods can be adapted to identify multiple jets.  For consistency of notation, we always use $R$ to refer to the adjustable jet radius parameter in each algorithm, which can be identified with the true jet radius only in the small $R$ limit.

\begin{itemize}
\item \textit{Jet function maximization} \cite{Georgi:2014zwa,Ge:2014ova,Bai:2014qca}.  Here, the jet finding strategy is to find the subset of particles in an event that maximizes a jet function $J(P_\mu)$, where $P_\mu$ is the four-vector of the candidate jet.\footnote{The name ``jet function'' and the symbol $J$ should not be confused with earlier usage in the context of factorization and resummation, e.g. \cite{Sterman:1986aj,Catani:1989ne,Korchemsky:1994jb}.}  The original paper \cite{Georgi:2014zwa} introduced a jet function appropriate for electron-positron collisions:
\be
\label{eq:Jorig}
J_{\rm orig}(P_\mu) = E - \frac{1}{R^2} \frac{m^2}{E},
\ee
where $E$ and $m$ are the total energy and mass of the candidate jet, and the original paper used the notation $R^2 \to 1/\beta$.  Maximizing this $J_{\rm orig}$ gives quasi-conical jets with nearly fixed radius $\simeq R$.  Since then, this algorithm has been adapted to collisions at the LHC \cite{Bai:2014qca} to yield nearly conical jets in the rapidity-azimuth plane.
\item \textit{1-jettiness minimization} \cite{Stewart:2010tn,Thaler:2010tr,Thaler:2011gf,Jouttenus:2013hs}.  This approach is based on finding a light-like jet axis $n = (1, \hat{n})$ that minimizes 1-jettiness $\Tau_1(n)$.  For electron-positron collisions, one possible choice of 1-jettiness measure is (see also \cite{Jouttenus:2013hs})
\be
\label{eq:tauorig}
\Tau_1(n) = \sum_{i \in {\rm event}} \min \left\{E_i, \frac{2 \, n\cdot p_i}{R^2} \right\},
\ee
where $p_i = (E_i, \vec{p}_i)$ is the four-vector of particle $i$, and one often sees the notation $R^2 \to 2 \rho$.  The minimum inside of $\Tau_1$ partitions the event into an unclustered (or ``beam'') region and a jet region, and after minimizing over $\hat{n}$, the jet region is exactly conical (for massless particles) with radius $\simeq R$.  This approach has been generalized to $N$-jettiness jet finding through the \textsc{XCone} jet algorithm \cite{Stewart:2015waa,Thaler:2015xaa}. 
\item \textit{Stable cone finding} \cite{Blazey:2000qt,Ellis:2001aa,Salam:2007xv}.   Algorithms like \textsc{SISCone} \cite{Salam:2007xv} search for all conical jet regions of radius $R$ which are stable, meaning that the total jet momentum is aligned with the jet axis.  This algorithm works equally well for opening angle as for rapidity-azimuth distance.  From the set of stable cones, one must then choose the desired jet(s).  By construction, the hardest jet after progressive removal (e.g.\ \textsc{SISCone-PR}  as described in \cite{fjmanual}) is a perfect cone.  While ``hardest'' typically refers to the jet with highest (transverse) momentum, we will find it interesting to consider the jet with largest $J$.   As shown in \cite{Ellis:2001aa}, stable cone finding can be viewed as an optimization problem, which is closely related to 1-jettiness minimization \cite{Thaler:2011gf}.  While historically there were issues with IRC safety of specific iterative cone algorithms (see discussion in \cite{Ellis:2007ib,Salam:2009jx}), those issues can be resolved using seedless methods (like in \textsc{SISCone}) or IRC safe seeds (like in \textsc{XCone}).
\end{itemize}

\section{A Meta Optimization Problem}
\label{sec:meta}

Since the three above approaches yield (nearly) conical jets and all are based on optimization, it is perhaps not surprising that, with suitable modifications, they can yield identical jets.  Here, we show this explicitly for cone jets of fixed opening angle, as relevant for electron-positron collisions.  We discuss obstructions to generalizing this to the LHC in the \Sec{sec:LHC}.

Consider the following meta function which depends both on a candidate jet $P_\mu = (E, \vec{P})$ as well as on an auxiliary light-like axis $n = (1, \hat{n})$:
\be
\label{eq:Mdef}
M(P_\mu,n) = E - \frac{2 \, n \cdot P}{R^2},
\ee
where $R$ will soon be identified with the same parameter in \Eq{eq:tauorig}.  Here, we work with massless final-state particles with $E_i = |\vec{p}_i|$; one can adapt this construction to massive particles by performing measurements in, say, the $p$-scheme or $E$-scheme where massive particles are replaced by massless proxies (see e.g.\ \cite{Salam:2001bd,Mateu:2012nk}). We claim that maximizing $M$ over all possible $P_\mu$ and $n$ simultaneously yields a jet function maximum, a 1-jettiness minimum, and a stable cone jet.

First, consider maximizing $M$ with respect to $n$, keeping the candidate jet $P_\mu$ fixed.  Introducing a Lagrange multiplier $\lambda$ to enforce unit norm 
\be
\label{eq:lagrange}
M(P_\mu,n) \rightarrow M(P_\mu,n) + \lambda (\hat{n}^2 - 1),
\ee
one can show that $M$ is maximized by
\be
n^{\rm opt} = \Biggl(1, \frac{\vec{P}}{|\vec{P}|} \Biggr).
\ee
Note that $\hat{n}$ lives on a smooth compact space (i.e.\ the surface of a unit sphere), so there are no edge conditions to check.  Plugging this back into \Eq{eq:Mdef}, we find
\be
M(P_\mu,n^{\rm opt}) = E - \frac{2 \bigl(E - |\vec{P}|\bigr)}{R^2} \equiv J(P_\mu).
\ee
Thus, maximizing $M$ over $\{P_\mu,n \}$ is equivalent to maximizing the jet function $J$ over $P_\mu$.  Since
\be
2\bigl(E - |\vec{P}|\bigr) \simeq \frac{m^2}{E}
\ee
for $m \ll E$, this jet function is very similar to the original definition in \Eq{eq:Jorig}.\footnote{\Ref{Ge:2014ova} introduced a generalization of \Eq{eq:Jorig} with an extra parameter $n$,
\be
J^{(n)} = E^n \left(1 - \frac{n}{R^2}\frac{m^2}{E^2} \right),
\ee
where $R^2 / n \to 1/\beta$ in the original notation.  Because $J^{(n)}$ is not linear in the jet kinematics, we know of no meta function that can directly handle this case.  That said, maximizing $J^{(n)}$ is equivalent to maximizing $\sqrt[n]{J^{(n)}}$, and $\sqrt[n]{J^{(n)}} \simeq J$ in the $m \ll E$ limit.}  
    Following the logic of \cite{Georgi:2014zwa}, a non-trivial $J$ maximum is guaranteed to exist since there are a finite number of $P_\mu$ partitions and a single particle has a larger $J$ value than the empty partition.

Next, consider maximizing $M$ over the candidate jet $P_\mu$, keeping $n$ fixed.  Crucially, $M$ is linear in the final state particles, so we can write $M$ as
\be
\label{eq:Mlinear}
M(P_\mu,n) = \sum_{i \in \rm jet}  E_i - \frac{2 \, n \cdot p_i}{R^2},
\ee
where ``jet'' refers to the set of particles contributing to $P_\mu$ (i.e.\ $\sum_{i \in \rm jet} E_i = E$, $\sum_{i \in \rm jet} \vec{p}_i = \vec{P}$).  Any final-state particle that contributes positively to $M$ will increase $M$'s value, so for the optimal jet $P^{\rm opt}_\mu$, we must have 
\be
\label{eq:Msumoverevent}
M(P^{\rm opt}_\mu,n) = \sum_{i \in {\rm event}} \max \left\{E_i - \frac{2 \, n \cdot p_i}{R^2} , 0\right \},
\ee
where now the sum runs over all particles in the event.  Rearranging this formula as
\be
E - M(P^{\rm opt}_\mu,n)= \sum_{i \in {\rm event}} \min \left\{E_i, \frac{2 \, n \cdot p_i}{R^2} \right\} \equiv \Tau_1(n),
\ee
we recover precisely \Eq{eq:tauorig}.  Thus, taking into account the minus sign, maximizing $M$ over $\{P_\mu,n \}$ is equivalent to minimizing 1-jettiness $\Tau_1$ over $n$.

Finally, we can maximize $M$ with respect to both $n$ and $P_\mu$ by combining these two analyses.  Since $M$ has a maximum, there must be an optimal subset $P^{\rm opt}_\mu$ (perhaps more than one in degenerate phase space configurations) and a corresponding optimal axis $n^{\rm opt}$.  Fixing $P^{\rm opt}_\mu$ and minimizing $\Tau_1(n)$ with respect to $n$ using the same Lagrange multiplier trick in \Eq{eq:lagrange}, the optimal axis must be 
\be
\label{eq:stablecone}
n^{\rm opt} = \Biggl(1, \frac{\vec{P}^{\rm opt}}{|\vec{P}^{\rm opt}|} \Biggr).
\ee
Now looking at \Eq{eq:Msumoverevent} with fixed $n^{\rm opt}$, the optimal subset $P_\mu^{\rm opt}$ must consist of all particles with
\be
\label{eq:Joptcondition}
\frac{2 \, n^{\rm opt} \cdot p_i}{E_i} < R^2.
\ee
This is precisely the condition for a stable cone.  To see why it is stable, note that \Eq{eq:stablecone} implies that the jet axis $n^{\rm opt}$ is aligned with the jet momentum $\vec{P}^{\rm opt}$.  To see why it is a cone, note that for massless particles with $|\vec{p}_i| = E_i$,
\be
\frac{2 \, n \cdot p_i}{E_i} = 2 - 2 \cos \theta_{\hat{n},i},
\ee 
where $\theta_{\hat{n},i}$ is the angle to the axis.  Thus, the condition in \Eq{eq:Joptcondition} specifies particles contained in a cone of half opening angle $R_{\rm true}$ around $n^{\rm opt}$ with the identification
\be
\label{eq:halfopen}
\sqrt{2 - 2 \cos R_{\rm true}} \equiv R,
\ee 
where $R_{\rm true} \simeq R$ at small radii.

We have therefore demonstrated that optimizing $M(P_\mu,n)$ is equivalent to optimizing the following two functions:
\begin{align}
\label{eq:Jnew}
J(P_\mu) &= E - \frac{2 \bigl(E - |\vec{P}|\bigr)}{R^2}, \\
\Tau_1(n) &= \sum_{i \in {\rm event}} \min \left\{E_i, \frac{2 \, n \cdot p_i}{R^2} \right\}.
\end{align}
Moreover, the optimal jet $P^{\rm opt}_\mu$ is specified by a stable cone centered on the axis $\hat{n}^{\rm opt} \propto \vec{P}^{\rm opt}$.  Of course, actually \emph{finding} the optimal configuration is a challenging computational problem.  In practice, one can make use of the converse statement that among all stable cones, the one that has the largest value of $J(P_\mu)$ is guaranteed to be the one that optimizes $M(P_\mu,n)$.  Therefore, one can run spherical \textsc{SISCone-PR} \cite{Salam:2007xv,fjmanual} to find all stable cone jets, and then simply select the jet with the largest value of $J(P_\mu)$.

\section{Two Particle Case Study}
\label{sec:twoparticle}

To better understand the above analysis, it is instructive to study the simplest case of two massless particles with equal energies $E/2$, separated by opening angle $\theta_{12}$.  For ease of notation, we define
\be
\sqrt{2 - 2 \cos \frac{\theta_{12}}{2}} \equiv \frac{R_{12}}{2}, \qquad \sqrt{2 - 2 \cos \theta_{12}} \equiv \widetilde{R}_{12},
\ee
such that $\theta_{12} \ge R_{12} \ge \widetilde{R}_{12}$, but all three agree in the small angle limit.

In the jet function approach using \Eq{eq:Jnew}, there are three candidate jets to consider:  particle 1 alone, particle 2 alone, or particles $1$ and $2$ together.  The corresponding jet function values are
\begin{align}
J(P_\mu^1) = J(P_\mu^2)    &= \frac{E}{2},\\
J(P_\mu^1 + P_\mu^2) &= E \left(1 - \frac{R_{12}^2}{4R^2} \right).
\end{align}
Choosing the maximum $J$ value, the condition for the two particles to be clustered into a single jet is
\be
\text{$J$ maximum:} \quad R_{12} < \sqrt{2} R.
\ee

In the 1-jettiness approach, it is straightforward to prove that the axis that minimizes $\Tau_1(n)$ must lie along the line between particles 1 and 2, such that
\be
\theta_{\hat{n},1} + \theta_{\hat{n},2} = \theta_{12}.
\ee
Furthermore, one can show that local minima of $\Tau_1(n)$ can only appear at three possible points:
\be
\label{eq:withmidpoint}
\theta_{\hat{n},1} = 0, \qquad \theta_{\hat{n},2} = 0, \qquad \theta_{\hat{n},1} = \theta_{\hat{n},2} = \frac{\theta_{12}}{2},
\ee
corresponding to the three candidate jet configurations above.  For $\widetilde{R}_{12} < R$ (note the tilde), only $\theta_{\hat{n},1} = \theta_{\hat{n},2}$ is a true local minimum, so the two particles are always clustered together.  For $R_{12} > 2 R$ (no tilde), only $\theta_{\hat{n},1} = 0$ and $\theta_{\hat{n},2} = 0$ are true local minima, so the two particles are never clustered together. For intermediate $R_{12}$ values, all three minima are present with corresponding 1-jettiness values of
\begin{align}
\Tau_1(\theta_{\hat{n},1} = 0) = \Tau_1(\theta_{\hat{n},2} = 0) &= \frac{E}{2},\\
\Tau_1(\theta_{\hat{n},1} = \theta_{\hat{n},2}) & =  \frac{E}{4} \frac{R_{12}^2}{R^2}.
\end{align}
Choosing the global $\Tau_1$ minimum, the particles will be clustered if
\be
\text{global $\Tau_1$ minimum:} \quad R_{12} < \sqrt{2} R,
\ee
yielding the same result as the jet function approach.   Alternatively, one might be content with only finding a local $\Tau_1$ minimum.  For example, \textsc{XCone} \cite{Stewart:2015waa,Thaler:2015xaa} searches for local $\Tau_1$ minima using IRC safe seeds based on $k_T$-style clustering \cite{Catani:1993hr,Ellis:1993tq}, leading to the clustering condition
\be
\text{$k_T$-seeded local $\Tau_1$ minimum:} \quad R_{12} < R.
\ee
With IRC safe seeds, this is a perfectly acceptable jet finding strategy from the point of view of perturbative QCD calculations.\footnote{Crucial to achieving IRC safety for \textsc{XCone}, the number of seeds is always equal to the desired number of jets $N$, independent of the final state particle multiplicity.}

Turning to the stable cone approach, the three local minima in $\Tau_1$ correspond precisely to the three possible stable cone configurations.  The default behavior of spherical \textsc{SISCone-PR} \cite{Salam:2007xv,fjmanual} is to progressively remove the stable cone with the largest energy, meaning that the two particles will always be clustered if they fall within a common cone:
\be
\text{$E$-ordered \textsc{SISCone-PR}:}  \quad R_{12} < 2R.
\ee
To mimic precisely the behavior of $J$ maximization or $\Tau_1$ minimization, though, one has to progressively remove the jet with the largest value of $J(P_\mu)$:
\be
\label{eq:Jordered}
\text{$J$-ordered \textsc{SISCone-PR}:}  \quad R_{12} < \sqrt{2} R.
\ee
Only for this version of stable cone finding does one recover the remarkable equivalence derived in \Sec{sec:meta}.  

As a historical note, the presence of local $\Tau_1$ minima led to many Tevatron-era discussions about cone jet algorithms \cite{Blazey:2000qt,Ellis:2001aa}, where a variant of $\Tau_1$ was known as the ``Snowmass potential''.  One issue is that of midpoint seeds, since if one only uses the location of the two particles themselves as seeds, then one misses out on a possible stable cone centered on their midpoint, as in \Eq{eq:withmidpoint}.  Since the location of $\Tau_1$ minima is an IRC safe property, this is the reason why seedless methods are preferred.   A related midpoint issue is that of disappearing minima, since in data, where the two particle directions are smeared out by showering and hadronization effects, the local $\Tau_1$ minimum at the midpoint disappears for $R_{12} \gtrsim R_{\rm sep} R$, where $R_{\rm sep} \simeq 1.3$  \cite{Ellis:1992qq}.  Disappearing minima do not present an issue for IRC safety, but they do lead to poor convergence of perturbation theory.  In that context, it is interesting to note that $1.3$ is comparable to the $\sqrt{2}$ factor in \Eq{eq:Jordered}.  This suggests that $J$-ordered cone finding may be more robust to showering and hadronization effects than $E$-ordered cone finding, since $J$-ordering aims to find global $\Tau_1$ minimum instead of just a local one.  We leave a detailed analysis of this possibility to future work.

\section{Differences at the LHC}
\label{sec:LHC}

For LHC applications, it is typically advantageous to work with transverse momenta and rapidity-azimuth distances.  Here, jet function maximization, 1-jettiness minimization, and stable cone finding are indeed inequivalent, which we now show.

Jet functions are based on optimizing the jet four-vector as a whole.  Therefore, we expect the jet function to be a function of the total transverse momentum $P_T$, total mass $m$, or the combination $E_T = \sqrt{P_T^2 + m^2}$.  For example, the chosen jet function in \cite{Bai:2014qca} was
\be
\label{eq:Jhad}
J_{\rm had}(P_\mu) = E_T - \frac{1}{R^2} \frac{m^2}{E_T},
\ee
where $R \to 1/ \sqrt{\beta}$ in the original notation.  As explained in \cite{Bai:2014qca}, while the jet region is nearly conical when using $J_{\rm had}$, the jet momentum direction is somewhat offset from the jet center at forward rapidities.

By contrast, 1-jettiness is based on optimizing a light-like axis, using a measure that is additive over the individual particles $i$.  As shown in \cite{Thaler:2011gf} (see also \cite{Stewart:2015waa,Thaler:2015xaa}), a cone algorithm can be defined via
\be
\label{eq:tau1had}
\Tau^{\rm had}_1(n) = \sum_{i \in {\rm event}} p_{Ti} \, \min \left\{1,  \frac{\Delta{R_{i,\hat{n}}}}{R} \right\}^\beta,
\ee
where $R$ is the jet radius and $\beta$ is an angular exponent (not to be confused with $\beta$ in the jet function literature).  For any value of $\beta$, the jet region is perfectly conical and the jet axis is located at the center of the jet region.  For the special value of $\beta = 2$, the jet axis is aligned with the $p_T$-weighted jet momentum (see \Eq{eq:stablecone1} below).  

At first glance, one might think that  $J_{\rm had}$ and $\Tau^{\rm had}_1$ could be massaged into a common meta function $M_{\rm had}$ by just making suitable $E \to p_T$ replacements.  However, there is a fundamental mismatch due to the fact that 
\be
\label{eq:mismatch}
\sum_i p_{Ti} \not= \Bigl(\sum_i p_i \Bigr)_T.
\ee
For this reason one cannot construct a suitable $M_{\rm had}$ that depends solely on $P_\mu$ and $n$, since one also needs information about the scalar $p_T$ sum within the jet.  Of course, one could find an $M_{\rm had}$ if one relaxes the requirements that $J_{\rm had}$ depends solely on $P_\mu$ or that $\Tau^{\rm had}_1$ can be expressed as a sum over $i$, though that goes somewhat against the original philosophies of those approaches.

\setlength{\tabcolsep}{0.2cm}
\begin{table*}
\begin{tabular}{rcccc|cc}
\hline \hline
 & $J_{E_T}$ & \textsc{XCone} $N =1$ & \textsc{SISCone-PR} & anti-$k_t$ & C/A &  $k_t$ \\
 \hline
$J_{E_T}$ &--- & 93\% & 96\% & 93\% & 81\% & 74\% \\
\textsc{XCone} $N =1$ &93\% & --- & 94\% & 96\% & 85\% & 77\% \\
\textsc{SISCone-PR} &96\% & 94\% & --- & 94\% & 82\% & 75\% \\
anti-$k_t$ & 93\% & 96\% & 94\% & --- & 84\% & 76\% \\
\hline
C/A & 81\% & 85\% & 82\% & 84\% & --- & 83\% \\
$k_t$ & 75\% & 78\% & 76\% & 76\% & 83\% & --- \\
\hline
\hline
\end{tabular}
\caption{Comparing the hardest jet found in $Z$ plus jets production at the LHC for $R = 0.4$.  Shown is the percentage of jets found by the column algorithm that match the row algorithm at the 3\% level, as measured by \Eq{eq:qualitymetric}.}
\label{fig:compare}
\end{table*}

\setlength{\tabcolsep}{0.2cm}
\begin{table*}
\begin{tabular}{rcccc|cc}
\hline \hline
 & $J_{E_T}$ & \textsc{XCone} $N =1$ & \textsc{SISCone-PR} & anti-$k_t$ & C/A &  $k_t$ \\
 \hline
$J_{E_T}$ &--- & 64\% & 72\% & 64\% & 45\% & 42\% \\
\textsc{XCone} $N=1$ &63\% & --- & 84\% & 91\% & 57\% & 48\% \\
\textsc{SISCone-PR} &71\% & 85\% & --- & 87\% & 53\% & 45\% \\
anti-$k_t$ & 63\% & 91\% & 87\% & --- & 56\% & 47\% \\
\hline
C/A & 44\% & 56\% & 53\% & 56\% & --- & 67\% \\
$k_t$ & 41\% & 48\% & 45\% & 47\% & 67\% & --- \\
\hline
\hline
\end{tabular}
\caption{Same as \Tab{fig:compare}, but for $R = 0.8$ where larger differences are expected.}
\label{fig:compare2}
\end{table*}

For stable cone finding, there are two different definitions in common use.  In \cite{Ellis:2001aa}, a stable cone is defined such that the jet axis aligns with the $E_T$-weighted centroid of its constituents.  For massless constituents, this is the same as minimizing $\Tau^{\rm had}_1$ in \Eq{eq:tau1had} with $\beta = 2$ \cite{Thaler:2011gf}, yielding a jet axis located at rapidity/azimuth location
\be
\label{eq:stablecone1}
y = \frac{\sum_{i \in {\rm jet}} y_i \, p_{Ti} }{\sum_{i \in {\rm jet}} p_{Ti}}, \qquad \phi = \frac{\sum_{i \in {\rm jet}} \phi_i \, p_{Ti} }{\sum_{i \in {\rm jet}} p_{Ti}}.
\ee
Alternatively, one can define a stable cone to have its jet axis aligned with the true jet three-momentum, which is the default behavior in \textsc{SISCone-PR} \cite{Salam:2007xv,fjmanual}.  For small radius jets, the distinction is small, but we know of no 1-jettiness measure (or jet function) that yields exactly stable cones by this second definition.

Despite the above discussion, the mismatch in \Eq{eq:mismatch} is a small effect at small $R$, with corrections that scale  like $R^2$ or $m^2/p_T^2$.  Moreover, these finite $R$ corrections are often subdominant to other effects present in practical jet algorithm implementations, such as the difference between local or global optimization or the treatment of jets at forward rapidities.

In typical situations, maximizing $J_{\rm had}$, minimizing $\Tau^{\rm had}_1$, and finding stable cones (by either definition) yield rather similar jets, which are also similar to anti-$k_t$ jets \cite{Cacciari:2008gp}.  This is shown in \Tab{fig:compare} for a $Z$ plus jets sample from \textsc{Pythia}~8.209 \cite{Sjostrand:2006za,Sjostrand:2007gs,Sjostrand:2014zea} at the 14 TeV LHC, looking at the hardest jet with $p_T > 50~\text{GeV}$ and $R = 0.4$ for all algorithms.  The jet function approach is implemented with the $J_{E_T}$ algorithm \cite{Bai:2014qca}, which finds jets ordered by \Eq{eq:Jhad} but we take the hardest jet by $p_T$.  The 1-jettiness approach is implemented with \textsc{XCone} \cite{Stewart:2015waa,Thaler:2015xaa}, using \Eq{eq:tau1had} with $\beta = 2$.\footnote{Note that \Eq{eq:tau1had} differs from the default \textsc{XCone} measure based on Lorentz dot products.  We use iterative one-pass minimization, which finds a local $\Tau_1^{\rm had}$ minimum from IRC safe seeds.}  The stable cone approach is implemented with \textsc{SISCone-PR} \cite{Salam:2007xv,fjmanual}, taking the hardest jet by $p_T$.  The anti-$k_t$ approach is implemented with \textsc{FastJet} \cite{Cacciari:2011ma}.  
For comparison, we also show results for the Cambridge/Aachen (C/A) \cite{Dokshitzer:1997in,Wobisch:1998wt,Wobisch:2000dk} and $k_t$ \cite{Catani:1993hr,Ellis:1993tq} algorithms.  As a similarity metric between the row and column jet algorithms in \Tab{fig:compare}, we look at the fraction of the scalar $p_T$ not contained in the shared constituents of the two algorithms: 
\be
\label{eq:qualitymetric}
\frac{\sum_{i \not\in \text{row} \cap \text{column}} p_{Ti}}{\sum_{i \in \text{row}} p_{Ti}}.
\ee
Around 95\% of the time, the four cone-like algorithms yield similar jet kinematics at the 3\% level, while the differences with C/A and $k_t$ are noticeably larger.  Because all of the algorithms considered are longitudinally boost invariant, the similarities/differences between algorithms are essentially independent of the jet rapidity.  In \Tab{fig:compare2}, we perform the same comparison for $R = 0.8$, where the differences are much larger as expected.

Note that the study in \Tab{fig:compare} involves events that typically only have one hard jet without substructure.  The four algorithms do exhibit larger variations when going to multiple jets, due to their differing treatment of jet overlap regions, though the algorithms still yield similar results when the jets are widely separated.  For jets with substructure, interesting differences can arise for subjets at wide angles (see further discussion in \cite{Bai:2014qca}).  Analogous to the discussion in \Sec{sec:twoparticle}, two particles with equal momentum sharing will be clustered together if:
\begin{equation}
\label{eq:Rsep12}
\Delta R_{12} \lesssim
\begin{cases}
R & \text{\textsc{XCone}, anti-$k_t$}, \\
\sqrt{2} R & \text{$J_{E_T}$}, \\
2 R &  \text{\textsc{SISCone-PR}},
\end{cases}
\end{equation}
where $2 \to R_{\rm sep}$ in the last line if there is sufficient smearing of the subjet directions.\footnote{The fact that the $J_{E_T}$ behavior is intermediate between anti-$k_t$ and \textsc{SISCone-PR} is also reflected in the NLO single jet inclusive cross section, where the $J_{E_T}$ result is very nearly the average of the anti-$k_t$ and \textsc{SISCone}-PR results \cite{Kaufmann:2014nda,Kaufmann:2015hma}.}  As in \Sec{sec:twoparticle}, we can make \textsc{SISCone-PR} behave more like the $J_{E_T}$ algorithm by switching from $p_T$-ordered to $J_{\rm had}$-ordered cones, and we can make \textsc{XCone} behave more like the $J_{E_T}$ algorithm by performing brute force global $\Tau_1$ minimization.  In cases like this where the four algorithms give different results, the desired behavior depends on the physics of interest.

\section{Conclusions}
\label{sec:conclude}

The idea of jet finding as an optimization problem has a long history (see also~\cite{Berger:2002jt,Angelini:2002et,Angelini:2004ac,Grigoriev:2003yc,Grigoriev:2003tn,Chekanov:2005cq,Lai:2008zp,Volobouev:2009rv}), so we find it remarkable that three seemingly different optimization strategies with different underlying philosophies correspond to the same meta optimization problem.  As we saw, the exact correspondence only holds for fixed-angle cone jets, whereas at the LHC, complications arise when projecting four-vectors to the rapidity/azimuth plane.  That said, the differences between the algorithms are small for small radius jets, and this correspondence helps explain why jet function maximization and 1-jettiness minimization yield such similar jets to stable cone jet algorithms.

There are different conclusions one might draw from this analysis.  To the extent that jet function maximization, 1-jettiness minimization, and stable cone finding yield similar jets (which are similar to anti-$k_t$), one should simply choose the jet algorithm with the best computational performance.  Alternatively, to the extent that the three algorithms have different underlying philosophies, one might try to exploit those differences to develop new jet finding techniques with qualitatively different behaviors or with better convergence in perturbative QCD calculations.  Ultimately, one should choose the jet finding approach that best suits the intended physics application.   In that spirit, this paper suggests that jet algorithms should be judged not by the elegance of their underlying philosophies but by the utility of their jet objects.

\begin{acknowledgments}
We thank Gavin Salam and Gregory Soyez for collaborating during the conceptual stages of this work and for subsequent suggestions.  We thank Steve Ellis and Andrew Larkoski for detailed comments on a draft of this manuscript.  We benefitted from conversations with Yang Bai, Yang-Ting Chien, Matteo Cacciari, Howard Georgi, Jakub Scholtz, Matthew Schwartz, Iain Stewart, Werner Vogelsang, and Bryan Webber.  J.T. is supported by the U.S. Department of Energy (DOE) under cooperative research agreement DE-SC-00012567, the DOE Early Career research program DE-SC-0006389, and by a Sloan Research Fellowship from the Alfred P. Sloan Foundation.
\end{acknowledgments}

\bibliography{SeparatedAtBirth}

\begin{thebibliography}{41}%
\makeatletter
\providecommand \@ifxundefined [1]{%
 \@ifx{#1\undefined}
}%
\providecommand \@ifnum [1]{%
 \ifnum #1\expandafter \@firstoftwo
 \else \expandafter \@secondoftwo
 \fi
}%
\providecommand \@ifx [1]{%
 \ifx #1\expandafter \@firstoftwo
 \else \expandafter \@secondoftwo
 \fi
}%
\providecommand \natexlab [1]{#1}%
\providecommand \enquote  [1]{``#1''}%
\providecommand \bibnamefont  [1]{#1}%
\providecommand \bibfnamefont [1]{#1}%
\providecommand \citenamefont [1]{#1}%
\providecommand \href@noop [0]{\@secondoftwo}%
\providecommand \href [0]{\begingroup \@sanitize@url \@href}%
\providecommand \@href[1]{\@@startlink{#1}\@@href}%
\providecommand \@@href[1]{\endgroup#1\@@endlink}%
\providecommand \@sanitize@url [0]{\catcode `\\12\catcode `\$12\catcode
  `\&12\catcode `\#12\catcode `\^12\catcode `\_12\catcode `\%12\relax}%
\providecommand \@@startlink[1]{}%
\providecommand \@@endlink[0]{}%
\providecommand \url  [0]{\begingroup\@sanitize@url \@url }%
\providecommand \@url [1]{\endgroup\@href {#1}{\urlprefix }}%
\providecommand \urlprefix  [0]{URL }%
\providecommand \Eprint [0]{\href }%
\providecommand \doibase [0]{http://dx.doi.org/}%
\providecommand \selectlanguage [0]{\@gobble}%
\providecommand \bibinfo  [0]{\@secondoftwo}%
\providecommand \bibfield  [0]{\@secondoftwo}%
\providecommand \translation [1]{[#1]}%
\providecommand \BibitemOpen [0]{}%
\providecommand \bibitemStop [0]{}%
\providecommand \bibitemNoStop [0]{.\EOS\space}%
\providecommand \EOS [0]{\spacefactor3000\relax}%
\providecommand \BibitemShut  [1]{\csname bibitem#1\endcsname}%
\let\auto@bib@innerbib\@empty
\bibitem [{\citenamefont {Ellis}\ \emph {et~al.}(2008)\citenamefont {Ellis},
  \citenamefont {Huston}, \citenamefont {Hatakeyama}, \citenamefont {Loch},\
  and\ \citenamefont {Tonnesmann}}]{Ellis:2007ib}%
  \BibitemOpen
  \bibfield  {author} {\bibinfo {author} {\bibfnamefont {S.}~\bibnamefont
  {Ellis}}, \bibinfo {author} {\bibfnamefont {J.}~\bibnamefont {Huston}},
  \bibinfo {author} {\bibfnamefont {K.}~\bibnamefont {Hatakeyama}}, \bibinfo
  {author} {\bibfnamefont {P.}~\bibnamefont {Loch}}, \ and\ \bibinfo {author}
  {\bibfnamefont {M.}~\bibnamefont {Tonnesmann}},\ }\href@noop {} {\bibfield
  {journal} {\bibinfo  {journal} {Prog.Part.Nucl.Phys.}\ }\textbf {\bibinfo
  {volume} {60}},\ \bibinfo {pages} {484} (\bibinfo {year} {2008})},\ \Eprint
  {http://arxiv.org/abs/0712.2447} {arXiv:0712.2447 [hep-ph]} \BibitemShut
  {NoStop}%
\bibitem [{\citenamefont {Salam}(2010)}]{Salam:2009jx}%
  \BibitemOpen
  \bibfield  {author} {\bibinfo {author} {\bibfnamefont {G.~P.}\ \bibnamefont
  {Salam}},\ }\href@noop {} {\bibfield  {journal} {\bibinfo  {journal}
  {Eur.Phys.J.}\ }\textbf {\bibinfo {volume} {C67}},\ \bibinfo {pages} {637}
  (\bibinfo {year} {2010})},\ \Eprint {http://arxiv.org/abs/0906.1833}
  {arXiv:0906.1833 [hep-ph]} \BibitemShut {NoStop}%
\bibitem [{\citenamefont {Georgi}(2014)}]{Georgi:2014zwa}%
  \BibitemOpen
  \bibfield  {author} {\bibinfo {author} {\bibfnamefont {H.}~\bibnamefont
  {Georgi}},\ }\href@noop {} {\  (\bibinfo {year} {2014})},\ \Eprint
  {http://arxiv.org/abs/1408.1161} {arXiv:1408.1161 [hep-ph]} \BibitemShut
  {NoStop}%
\bibitem [{\citenamefont {Ge}(2015)}]{Ge:2014ova}%
  \BibitemOpen
  \bibfield  {author} {\bibinfo {author} {\bibfnamefont {S.-F.}\ \bibnamefont
  {Ge}},\ }\href {\doibase 10.1007/JHEP05(2015)066} {\bibfield  {journal}
  {\bibinfo  {journal} {JHEP}\ }\textbf {\bibinfo {volume} {05}},\ \bibinfo
  {pages} {066} (\bibinfo {year} {2015})},\ \Eprint
  {http://arxiv.org/abs/1408.3823} {arXiv:1408.3823 [hep-ph]} \BibitemShut
  {NoStop}%
\bibitem [{\citenamefont {Bai}\ \emph {et~al.}(2015)\citenamefont {Bai},
  \citenamefont {Han},\ and\ \citenamefont {Lu}}]{Bai:2014qca}%
  \BibitemOpen
  \bibfield  {author} {\bibinfo {author} {\bibfnamefont {Y.}~\bibnamefont
  {Bai}}, \bibinfo {author} {\bibfnamefont {Z.}~\bibnamefont {Han}}, \ and\
  \bibinfo {author} {\bibfnamefont {R.}~\bibnamefont {Lu}},\ }\href@noop {}
  {\bibfield  {journal} {\bibinfo  {journal} {JHEP}\ }\textbf {\bibinfo
  {volume} {1503}},\ \bibinfo {pages} {102} (\bibinfo {year} {2015})},\ \Eprint
  {http://arxiv.org/abs/1411.3705} {arXiv:1411.3705 [hep-ph]} \BibitemShut
  {NoStop}%
\bibitem [{\citenamefont {Stewart}\ \emph {et~al.}(2010)\citenamefont
  {Stewart}, \citenamefont {Tackmann},\ and\ \citenamefont
  {Waalewijn}}]{Stewart:2010tn}%
  \BibitemOpen
  \bibfield  {author} {\bibinfo {author} {\bibfnamefont {I.~W.}\ \bibnamefont
  {Stewart}}, \bibinfo {author} {\bibfnamefont {F.~J.}\ \bibnamefont
  {Tackmann}}, \ and\ \bibinfo {author} {\bibfnamefont {W.~J.}\ \bibnamefont
  {Waalewijn}},\ }\href {\doibase 10.1103/PhysRevLett.105.092002} {\bibfield
  {journal} {\bibinfo  {journal} {Phys.Rev.Lett.}\ }\textbf {\bibinfo {volume}
  {105}},\ \bibinfo {pages} {092002} (\bibinfo {year} {2010})},\ \Eprint
  {http://arxiv.org/abs/1004.2489} {arXiv:1004.2489 [hep-ph]} \BibitemShut
  {NoStop}%
\bibitem [{\citenamefont {Thaler}\ and\ \citenamefont
  {Van~Tilburg}(2011)}]{Thaler:2010tr}%
  \BibitemOpen
  \bibfield  {author} {\bibinfo {author} {\bibfnamefont {J.}~\bibnamefont
  {Thaler}}\ and\ \bibinfo {author} {\bibfnamefont {K.}~\bibnamefont
  {Van~Tilburg}},\ }\href@noop {} {\bibfield  {journal} {\bibinfo  {journal}
  {JHEP}\ }\textbf {\bibinfo {volume} {1103}},\ \bibinfo {pages} {015}
  (\bibinfo {year} {2011})},\ \Eprint {http://arxiv.org/abs/1011.2268}
  {arXiv:1011.2268 [hep-ph]} \BibitemShut {NoStop}%
\bibitem [{\citenamefont {Thaler}\ and\ \citenamefont
  {Van~Tilburg}(2012)}]{Thaler:2011gf}%
  \BibitemOpen
  \bibfield  {author} {\bibinfo {author} {\bibfnamefont {J.}~\bibnamefont
  {Thaler}}\ and\ \bibinfo {author} {\bibfnamefont {K.}~\bibnamefont
  {Van~Tilburg}},\ }\href@noop {} {\bibfield  {journal} {\bibinfo  {journal}
  {JHEP}\ }\textbf {\bibinfo {volume} {1202}},\ \bibinfo {pages} {093}
  (\bibinfo {year} {2012})},\ \Eprint {http://arxiv.org/abs/1108.2701}
  {arXiv:1108.2701 [hep-ph]} \BibitemShut {NoStop}%
\bibitem [{\citenamefont {Jouttenus}\ \emph {et~al.}(2013)\citenamefont
  {Jouttenus}, \citenamefont {Stewart}, \citenamefont {Tackmann},\ and\
  \citenamefont {Waalewijn}}]{Jouttenus:2013hs}%
  \BibitemOpen
  \bibfield  {author} {\bibinfo {author} {\bibfnamefont {T.~T.}\ \bibnamefont
  {Jouttenus}}, \bibinfo {author} {\bibfnamefont {I.~W.}\ \bibnamefont
  {Stewart}}, \bibinfo {author} {\bibfnamefont {F.~J.}\ \bibnamefont
  {Tackmann}}, \ and\ \bibinfo {author} {\bibfnamefont {W.~J.}\ \bibnamefont
  {Waalewijn}},\ }\href@noop {} {\bibfield  {journal} {\bibinfo  {journal}
  {Phys.Rev.}\ }\textbf {\bibinfo {volume} {D88}},\ \bibinfo {pages} {054031}
  (\bibinfo {year} {2013})},\ \Eprint {http://arxiv.org/abs/1302.0846}
  {arXiv:1302.0846 [hep-ph]} \BibitemShut {NoStop}%
\bibitem [{\citenamefont {Blazey}\ \emph {et~al.}(2000)\citenamefont {Blazey},
  \citenamefont {Dittmann}, \citenamefont {Ellis}, \citenamefont {Elvira},
  \citenamefont {Frame} \emph {et~al.}}]{Blazey:2000qt}%
  \BibitemOpen
  \bibfield  {author} {\bibinfo {author} {\bibfnamefont {G.~C.}\ \bibnamefont
  {Blazey}}, \bibinfo {author} {\bibfnamefont {J.~R.}\ \bibnamefont
  {Dittmann}}, \bibinfo {author} {\bibfnamefont {S.~D.}\ \bibnamefont {Ellis}},
  \bibinfo {author} {\bibfnamefont {V.~D.}\ \bibnamefont {Elvira}}, \bibinfo
  {author} {\bibfnamefont {K.}~\bibnamefont {Frame}},  \emph {et~al.},\
  }\href@noop {} {\ ,\ \bibinfo {pages} {47} (\bibinfo {year} {2000})},\
  \Eprint {http://arxiv.org/abs/hep-ex/0005012} {arXiv:hep-ex/0005012 [hep-ex]}
  \BibitemShut {NoStop}%
\bibitem [{\citenamefont {Ellis}\ \emph {et~al.}(2001)\citenamefont {Ellis},
  \citenamefont {Huston},\ and\ \citenamefont {Tonnesmann}}]{Ellis:2001aa}%
  \BibitemOpen
  \bibfield  {author} {\bibinfo {author} {\bibfnamefont {S.}~\bibnamefont
  {Ellis}}, \bibinfo {author} {\bibfnamefont {J.}~\bibnamefont {Huston}}, \
  and\ \bibinfo {author} {\bibfnamefont {M.}~\bibnamefont {Tonnesmann}},\
  }\href@noop {} {\ ,\ \bibinfo {pages} {P513} (\bibinfo {year} {2001})},\
  \Eprint {http://arxiv.org/abs/hep-ph/0111434} {arXiv:hep-ph/0111434 [hep-ph]}
  \BibitemShut {NoStop}%
\bibitem [{\citenamefont {Salam}\ and\ \citenamefont
  {Soyez}(2007)}]{Salam:2007xv}%
  \BibitemOpen
  \bibfield  {author} {\bibinfo {author} {\bibfnamefont {G.~P.}\ \bibnamefont
  {Salam}}\ and\ \bibinfo {author} {\bibfnamefont {G.}~\bibnamefont {Soyez}},\
  }\href@noop {} {\bibfield  {journal} {\bibinfo  {journal} {JHEP}\ }\textbf
  {\bibinfo {volume} {0705}},\ \bibinfo {pages} {086} (\bibinfo {year}
  {2007})},\ \Eprint {http://arxiv.org/abs/0704.0292} {arXiv:0704.0292
  [hep-ph]} \BibitemShut {NoStop}%
\bibitem [{\citenamefont {Sterman}(1987)}]{Sterman:1986aj}%
  \BibitemOpen
  \bibfield  {author} {\bibinfo {author} {\bibfnamefont {G.~F.}\ \bibnamefont
  {Sterman}},\ }\href {\doibase 10.1016/0550-3213(87)90258-6} {\bibfield
  {journal} {\bibinfo  {journal} {Nucl.Phys.}\ }\textbf {\bibinfo {volume}
  {B281}},\ \bibinfo {pages} {310} (\bibinfo {year} {1987})}\BibitemShut
  {NoStop}%
\bibitem [{\citenamefont {Catani}\ and\ \citenamefont
  {Trentadue}(1989)}]{Catani:1989ne}%
  \BibitemOpen
  \bibfield  {author} {\bibinfo {author} {\bibfnamefont {S.}~\bibnamefont
  {Catani}}\ and\ \bibinfo {author} {\bibfnamefont {L.}~\bibnamefont
  {Trentadue}},\ }\href {\doibase 10.1016/0550-3213(89)90273-3} {\bibfield
  {journal} {\bibinfo  {journal} {Nucl.Phys.}\ }\textbf {\bibinfo {volume}
  {B327}},\ \bibinfo {pages} {323} (\bibinfo {year} {1989})}\BibitemShut
  {NoStop}%
\bibitem [{\citenamefont {Korchemsky}\ and\ \citenamefont
  {Sterman}(1994)}]{Korchemsky:1994jb}%
  \BibitemOpen
  \bibfield  {author} {\bibinfo {author} {\bibfnamefont {G.~P.}\ \bibnamefont
  {Korchemsky}}\ and\ \bibinfo {author} {\bibfnamefont {G.~F.}\ \bibnamefont
  {Sterman}},\ }\href {\doibase 10.1016/0370-2693(94)91304-8} {\bibfield
  {journal} {\bibinfo  {journal} {Phys.Lett.}\ }\textbf {\bibinfo {volume}
  {B340}},\ \bibinfo {pages} {96} (\bibinfo {year} {1994})},\ \Eprint
  {http://arxiv.org/abs/hep-ph/9407344} {arXiv:hep-ph/9407344 [hep-ph]}
  \BibitemShut {NoStop}%
\bibitem [{\citenamefont {Stewart}\ \emph {et~al.}(2015)\citenamefont
  {Stewart}, \citenamefont {Tackmann}, \citenamefont {Thaler}, \citenamefont
  {Vermilion},\ and\ \citenamefont {Wilkason}}]{Stewart:2015waa}%
  \BibitemOpen
  \bibfield  {author} {\bibinfo {author} {\bibfnamefont {I.~W.}\ \bibnamefont
  {Stewart}}, \bibinfo {author} {\bibfnamefont {F.~J.}\ \bibnamefont
  {Tackmann}}, \bibinfo {author} {\bibfnamefont {J.}~\bibnamefont {Thaler}},
  \bibinfo {author} {\bibfnamefont {C.~K.}\ \bibnamefont {Vermilion}}, \ and\
  \bibinfo {author} {\bibfnamefont {T.~F.}\ \bibnamefont {Wilkason}},\
  }\href@noop {} {\  (\bibinfo {year} {2015})},\ \Eprint
  {http://arxiv.org/abs/1508.01516} {arXiv:1508.01516 [hep-ph]} \BibitemShut
  {NoStop}%
\bibitem [{\citenamefont {Thaler}\ and\ \citenamefont
  {Wilkason}(2015)}]{Thaler:2015xaa}%
  \BibitemOpen
  \bibfield  {author} {\bibinfo {author} {\bibfnamefont {J.}~\bibnamefont
  {Thaler}}\ and\ \bibinfo {author} {\bibfnamefont {T.~F.}\ \bibnamefont
  {Wilkason}},\ }\href@noop {} {\  (\bibinfo {year} {2015})},\ \Eprint
  {http://arxiv.org/abs/1508.01518} {arXiv:1508.01518 [hep-ph]} \BibitemShut
  {NoStop}%
\bibitem [{fjm()}]{fjmanual}%
  \BibitemOpen
  \href@noop {} {\enquote {\bibinfo {title}
  {http://fastjet.fr/repo/fastjet-doc-3.1.2.pdf},}\ }\BibitemShut {NoStop}%
\bibitem [{\citenamefont {Salam}\ and\ \citenamefont
  {Wicke}(2001)}]{Salam:2001bd}%
  \BibitemOpen
  \bibfield  {author} {\bibinfo {author} {\bibfnamefont {G.~P.}\ \bibnamefont
  {Salam}}\ and\ \bibinfo {author} {\bibfnamefont {D.}~\bibnamefont {Wicke}},\
  }\href@noop {} {\bibfield  {journal} {\bibinfo  {journal} {JHEP}\ }\textbf
  {\bibinfo {volume} {05}},\ \bibinfo {pages} {061} (\bibinfo {year} {2001})},\
  \Eprint {http://arxiv.org/abs/hep-ph/0102343} {arXiv:hep-ph/0102343}
  \BibitemShut {NoStop}%
\bibitem [{\citenamefont {Mateu}\ \emph {et~al.}(2013)\citenamefont {Mateu},
  \citenamefont {Stewart},\ and\ \citenamefont {Thaler}}]{Mateu:2012nk}%
  \BibitemOpen
  \bibfield  {author} {\bibinfo {author} {\bibfnamefont {V.}~\bibnamefont
  {Mateu}}, \bibinfo {author} {\bibfnamefont {I.~W.}\ \bibnamefont {Stewart}},
  \ and\ \bibinfo {author} {\bibfnamefont {J.}~\bibnamefont {Thaler}},\ }\href
  {\doibase 10.1103/PhysRevD.87.014025} {\bibfield  {journal} {\bibinfo
  {journal} {Phys.Rev.}\ }\textbf {\bibinfo {volume} {D87}},\ \bibinfo {pages}
  {014025} (\bibinfo {year} {2013})},\ \Eprint {http://arxiv.org/abs/1209.3781}
  {arXiv:1209.3781 [hep-ph]} \BibitemShut {NoStop}%
\bibitem [{\citenamefont {Catani}\ \emph {et~al.}(1993)\citenamefont {Catani},
  \citenamefont {Dokshitzer}, \citenamefont {Seymour},\ and\ \citenamefont
  {Webber}}]{Catani:1993hr}%
  \BibitemOpen
  \bibfield  {author} {\bibinfo {author} {\bibfnamefont {S.}~\bibnamefont
  {Catani}}, \bibinfo {author} {\bibfnamefont {Y.~L.}\ \bibnamefont
  {Dokshitzer}}, \bibinfo {author} {\bibfnamefont {M.}~\bibnamefont {Seymour}},
  \ and\ \bibinfo {author} {\bibfnamefont {B.}~\bibnamefont {Webber}},\
  }\href@noop {} {\bibfield  {journal} {\bibinfo  {journal} {Nucl.Phys.}\
  }\textbf {\bibinfo {volume} {B406}},\ \bibinfo {pages} {187} (\bibinfo {year}
  {1993})}\BibitemShut {NoStop}%
\bibitem [{\citenamefont {Ellis}\ and\ \citenamefont
  {Soper}(1993)}]{Ellis:1993tq}%
  \BibitemOpen
  \bibfield  {author} {\bibinfo {author} {\bibfnamefont {S.~D.}\ \bibnamefont
  {Ellis}}\ and\ \bibinfo {author} {\bibfnamefont {D.~E.}\ \bibnamefont
  {Soper}},\ }\href@noop {} {\bibfield  {journal} {\bibinfo  {journal}
  {Phys.Rev.}\ }\textbf {\bibinfo {volume} {D48}},\ \bibinfo {pages} {3160}
  (\bibinfo {year} {1993})},\ \Eprint {http://arxiv.org/abs/hep-ph/9305266}
  {arXiv:hep-ph/9305266 [hep-ph]} \BibitemShut {NoStop}%
\bibitem [{\citenamefont {Ellis}\ \emph {et~al.}(1992)\citenamefont {Ellis},
  \citenamefont {Kunszt},\ and\ \citenamefont {Soper}}]{Ellis:1992qq}%
  \BibitemOpen
  \bibfield  {author} {\bibinfo {author} {\bibfnamefont {S.~D.}\ \bibnamefont
  {Ellis}}, \bibinfo {author} {\bibfnamefont {Z.}~\bibnamefont {Kunszt}}, \
  and\ \bibinfo {author} {\bibfnamefont {D.~E.}\ \bibnamefont {Soper}},\ }\href
  {\doibase 10.1103/PhysRevLett.69.3615} {\bibfield  {journal} {\bibinfo
  {journal} {Phys.Rev.Lett.}\ }\textbf {\bibinfo {volume} {69}},\ \bibinfo
  {pages} {3615} (\bibinfo {year} {1992})},\ \Eprint
  {http://arxiv.org/abs/hep-ph/9208249} {arXiv:hep-ph/9208249 [hep-ph]}
  \BibitemShut {NoStop}%
\bibitem [{\citenamefont {Cacciari}\ \emph {et~al.}(2008)\citenamefont
  {Cacciari}, \citenamefont {Salam},\ and\ \citenamefont
  {Soyez}}]{Cacciari:2008gp}%
  \BibitemOpen
  \bibfield  {author} {\bibinfo {author} {\bibfnamefont {M.}~\bibnamefont
  {Cacciari}}, \bibinfo {author} {\bibfnamefont {G.~P.}\ \bibnamefont {Salam}},
  \ and\ \bibinfo {author} {\bibfnamefont {G.}~\bibnamefont {Soyez}},\
  }\href@noop {} {\bibfield  {journal} {\bibinfo  {journal} {JHEP}\ }\textbf
  {\bibinfo {volume} {0804}},\ \bibinfo {pages} {063} (\bibinfo {year}
  {2008})},\ \Eprint {http://arxiv.org/abs/0802.1189} {arXiv:0802.1189
  [hep-ph]} \BibitemShut {NoStop}%
\bibitem [{\citenamefont {Sjostrand}\ \emph {et~al.}(2006)\citenamefont
  {Sjostrand}, \citenamefont {Mrenna},\ and\ \citenamefont
  {Skands}}]{Sjostrand:2006za}%
  \BibitemOpen
  \bibfield  {author} {\bibinfo {author} {\bibfnamefont {T.}~\bibnamefont
  {Sjostrand}}, \bibinfo {author} {\bibfnamefont {S.}~\bibnamefont {Mrenna}}, \
  and\ \bibinfo {author} {\bibfnamefont {P.~Z.}\ \bibnamefont {Skands}},\
  }\href@noop {} {\bibfield  {journal} {\bibinfo  {journal} {JHEP}\ }\textbf
  {\bibinfo {volume} {0605}},\ \bibinfo {pages} {026} (\bibinfo {year}
  {2006})},\ \Eprint {http://arxiv.org/abs/hep-ph/0603175}
  {arXiv:hep-ph/0603175 [hep-ph]} \BibitemShut {NoStop}%
\bibitem [{\citenamefont {Sjostrand}\ \emph {et~al.}(2008)\citenamefont
  {Sjostrand}, \citenamefont {Mrenna},\ and\ \citenamefont
  {Skands}}]{Sjostrand:2007gs}%
  \BibitemOpen
  \bibfield  {author} {\bibinfo {author} {\bibfnamefont {T.}~\bibnamefont
  {Sjostrand}}, \bibinfo {author} {\bibfnamefont {S.}~\bibnamefont {Mrenna}}, \
  and\ \bibinfo {author} {\bibfnamefont {P.~Z.}\ \bibnamefont {Skands}},\
  }\href@noop {} {\bibfield  {journal} {\bibinfo  {journal}
  {Comput.Phys.Commun.}\ }\textbf {\bibinfo {volume} {178}},\ \bibinfo {pages}
  {852} (\bibinfo {year} {2008})},\ \Eprint {http://arxiv.org/abs/0710.3820}
  {arXiv:0710.3820 [hep-ph]} \BibitemShut {NoStop}%
\bibitem [{\citenamefont {Sjöstrand}\ \emph {et~al.}(2015)\citenamefont
  {Sjöstrand}, \citenamefont {Ask}, \citenamefont {Christiansen},
  \citenamefont {Corke}, \citenamefont {Desai} \emph
  {et~al.}}]{Sjostrand:2014zea}%
  \BibitemOpen
  \bibfield  {author} {\bibinfo {author} {\bibfnamefont {T.}~\bibnamefont
  {Sjöstrand}}, \bibinfo {author} {\bibfnamefont {S.}~\bibnamefont {Ask}},
  \bibinfo {author} {\bibfnamefont {J.~R.}\ \bibnamefont {Christiansen}},
  \bibinfo {author} {\bibfnamefont {R.}~\bibnamefont {Corke}}, \bibinfo
  {author} {\bibfnamefont {N.}~\bibnamefont {Desai}},  \emph {et~al.},\ }\href
  {\doibase 10.1016/j.cpc.2015.01.024} {\bibfield  {journal} {\bibinfo
  {journal} {Comput.Phys.Commun.}\ }\textbf {\bibinfo {volume} {191}},\
  \bibinfo {pages} {159} (\bibinfo {year} {2015})},\ \Eprint
  {http://arxiv.org/abs/1410.3012} {arXiv:1410.3012 [hep-ph]} \BibitemShut
  {NoStop}%
\bibitem [{\citenamefont {Cacciari}\ \emph {et~al.}(2012)\citenamefont
  {Cacciari}, \citenamefont {Salam},\ and\ \citenamefont
  {Soyez}}]{Cacciari:2011ma}%
  \BibitemOpen
  \bibfield  {author} {\bibinfo {author} {\bibfnamefont {M.}~\bibnamefont
  {Cacciari}}, \bibinfo {author} {\bibfnamefont {G.~P.}\ \bibnamefont {Salam}},
  \ and\ \bibinfo {author} {\bibfnamefont {G.}~\bibnamefont {Soyez}},\
  }\href@noop {} {\bibfield  {journal} {\bibinfo  {journal} {Eur.Phys.J.}\
  }\textbf {\bibinfo {volume} {C72}},\ \bibinfo {pages} {1896} (\bibinfo {year}
  {2012})},\ \Eprint {http://arxiv.org/abs/1111.6097} {arXiv:1111.6097
  [hep-ph]} \BibitemShut {NoStop}%
\bibitem [{\citenamefont {Dokshitzer}\ \emph {et~al.}(1997)\citenamefont
  {Dokshitzer}, \citenamefont {Leder}, \citenamefont {Moretti},\ and\
  \citenamefont {Webber}}]{Dokshitzer:1997in}%
  \BibitemOpen
  \bibfield  {author} {\bibinfo {author} {\bibfnamefont {Y.~L.}\ \bibnamefont
  {Dokshitzer}}, \bibinfo {author} {\bibfnamefont {G.}~\bibnamefont {Leder}},
  \bibinfo {author} {\bibfnamefont {S.}~\bibnamefont {Moretti}}, \ and\
  \bibinfo {author} {\bibfnamefont {B.}~\bibnamefont {Webber}},\ }\href@noop {}
  {\bibfield  {journal} {\bibinfo  {journal} {JHEP}\ }\textbf {\bibinfo
  {volume} {9708}},\ \bibinfo {pages} {001} (\bibinfo {year} {1997})},\ \Eprint
  {http://arxiv.org/abs/hep-ph/9707323} {arXiv:hep-ph/9707323 [hep-ph]}
  \BibitemShut {NoStop}%
\bibitem [{\citenamefont {Wobisch}\ and\ \citenamefont
  {Wengler}(1998)}]{Wobisch:1998wt}%
  \BibitemOpen
  \bibfield  {author} {\bibinfo {author} {\bibfnamefont {M.}~\bibnamefont
  {Wobisch}}\ and\ \bibinfo {author} {\bibfnamefont {T.}~\bibnamefont
  {Wengler}},\ }\href@noop {} {\  (\bibinfo {year} {1998})},\ \Eprint
  {http://arxiv.org/abs/hep-ph/9907280} {arXiv:hep-ph/9907280 [hep-ph]}
  \BibitemShut {NoStop}%
\bibitem [{\citenamefont {Wobisch}(2000)}]{Wobisch:2000dk}%
  \BibitemOpen
  \bibfield  {author} {\bibinfo {author} {\bibfnamefont {M.}~\bibnamefont
  {Wobisch}},\ }\href@noop {} {\bibfield  {journal} {\bibinfo  {journal}
  {DESY-THESIS-2000-049}\ } (\bibinfo {year} {2000})}\BibitemShut {NoStop}%
\bibitem [{\citenamefont {Kaufmann}\ \emph
  {et~al.}(2015{\natexlab{a}})\citenamefont {Kaufmann}, \citenamefont
  {Mukherjee},\ and\ \citenamefont {Vogelsang}}]{Kaufmann:2014nda}%
  \BibitemOpen
  \bibfield  {author} {\bibinfo {author} {\bibfnamefont {T.}~\bibnamefont
  {Kaufmann}}, \bibinfo {author} {\bibfnamefont {A.}~\bibnamefont {Mukherjee}},
  \ and\ \bibinfo {author} {\bibfnamefont {W.}~\bibnamefont {Vogelsang}},\
  }\href {\doibase 10.1103/PhysRevD.91.034001} {\bibfield  {journal} {\bibinfo
  {journal} {Phys. Rev.}\ }\textbf {\bibinfo {volume} {D91}},\ \bibinfo {pages}
  {034001} (\bibinfo {year} {2015}{\natexlab{a}})},\ \Eprint
  {http://arxiv.org/abs/1412.0298} {arXiv:1412.0298 [hep-ph]} \BibitemShut
  {NoStop}%
\bibitem [{\citenamefont {Kaufmann}\ \emph
  {et~al.}(2015{\natexlab{b}})\citenamefont {Kaufmann}, \citenamefont
  {Mukherjee},\ and\ \citenamefont {Vogelsang}}]{Kaufmann:2015hma}%
  \BibitemOpen
  \bibfield  {author} {\bibinfo {author} {\bibfnamefont {T.}~\bibnamefont
  {Kaufmann}}, \bibinfo {author} {\bibfnamefont {A.}~\bibnamefont {Mukherjee}},
  \ and\ \bibinfo {author} {\bibfnamefont {W.}~\bibnamefont {Vogelsang}},\
  }\href@noop {} {\  (\bibinfo {year} {2015}{\natexlab{b}})},\ \Eprint
  {http://arxiv.org/abs/1506.01415} {arXiv:1506.01415 [hep-ph]} \BibitemShut
  {NoStop}%
\bibitem [{\citenamefont {Berger}\ \emph {et~al.}(2001)\citenamefont {Berger},
  \citenamefont {Berger}, \citenamefont {Bhat}, \citenamefont {Butterworth},
  \citenamefont {Ellis} \emph {et~al.}}]{Berger:2002jt}%
  \BibitemOpen
  \bibfield  {author} {\bibinfo {author} {\bibfnamefont {C.}~\bibnamefont
  {Berger}}, \bibinfo {author} {\bibfnamefont {E.~L.}\ \bibnamefont {Berger}},
  \bibinfo {author} {\bibfnamefont {P.}~\bibnamefont {Bhat}}, \bibinfo {author}
  {\bibfnamefont {J.}~\bibnamefont {Butterworth}}, \bibinfo {author}
  {\bibfnamefont {S.}~\bibnamefont {Ellis}},  \emph {et~al.},\ }\href@noop {}
  {\ ,\ \bibinfo {pages} {P512} (\bibinfo {year} {2001})},\ \Eprint
  {http://arxiv.org/abs/hep-ph/0202207} {arXiv:hep-ph/0202207 [hep-ph]}
  \BibitemShut {NoStop}%
\bibitem [{\citenamefont {Angelini}\ \emph {et~al.}(2002)\citenamefont
  {Angelini}, \citenamefont {De~Felice}, \citenamefont {Maggi}, \citenamefont
  {Nardulli}, \citenamefont {Nitti} \emph {et~al.}}]{Angelini:2002et}%
  \BibitemOpen
  \bibfield  {author} {\bibinfo {author} {\bibfnamefont {L.}~\bibnamefont
  {Angelini}}, \bibinfo {author} {\bibfnamefont {P.}~\bibnamefont {De~Felice}},
  \bibinfo {author} {\bibfnamefont {M.}~\bibnamefont {Maggi}}, \bibinfo
  {author} {\bibfnamefont {G.}~\bibnamefont {Nardulli}}, \bibinfo {author}
  {\bibfnamefont {L.}~\bibnamefont {Nitti}},  \emph {et~al.},\ }\href@noop {}
  {\bibfield  {journal} {\bibinfo  {journal} {Phys.Lett.}\ }\textbf {\bibinfo
  {volume} {B545}},\ \bibinfo {pages} {315} (\bibinfo {year} {2002})},\ \Eprint
  {http://arxiv.org/abs/hep-ph/0207032} {arXiv:hep-ph/0207032 [hep-ph]}
  \BibitemShut {NoStop}%
\bibitem [{\citenamefont {Angelini}\ \emph {et~al.}(2004)\citenamefont
  {Angelini}, \citenamefont {Nardulli}, \citenamefont {Nitti}, \citenamefont
  {Pellicoro}, \citenamefont {Perrino} \emph {et~al.}}]{Angelini:2004ac}%
  \BibitemOpen
  \bibfield  {author} {\bibinfo {author} {\bibfnamefont {L.}~\bibnamefont
  {Angelini}}, \bibinfo {author} {\bibfnamefont {G.}~\bibnamefont {Nardulli}},
  \bibinfo {author} {\bibfnamefont {L.}~\bibnamefont {Nitti}}, \bibinfo
  {author} {\bibfnamefont {M.}~\bibnamefont {Pellicoro}}, \bibinfo {author}
  {\bibfnamefont {D.}~\bibnamefont {Perrino}},  \emph {et~al.},\ }\href@noop {}
  {\bibfield  {journal} {\bibinfo  {journal} {Phys.Lett.}\ }\textbf {\bibinfo
  {volume} {B601}},\ \bibinfo {pages} {56} (\bibinfo {year} {2004})},\ \Eprint
  {http://arxiv.org/abs/hep-ph/0407214} {arXiv:hep-ph/0407214 [hep-ph]}
  \BibitemShut {NoStop}%
\bibitem [{\citenamefont {Grigoriev}\ \emph
  {et~al.}(2003{\natexlab{a}})\citenamefont {Grigoriev}, \citenamefont
  {Jankowski},\ and\ \citenamefont {Tkachov}}]{Grigoriev:2003yc}%
  \BibitemOpen
  \bibfield  {author} {\bibinfo {author} {\bibfnamefont {D.}~\bibnamefont
  {Grigoriev}}, \bibinfo {author} {\bibfnamefont {E.}~\bibnamefont
  {Jankowski}}, \ and\ \bibinfo {author} {\bibfnamefont {F.}~\bibnamefont
  {Tkachov}},\ }\href@noop {} {\bibfield  {journal} {\bibinfo  {journal}
  {Phys.Rev.Lett.}\ }\textbf {\bibinfo {volume} {91}},\ \bibinfo {pages}
  {061801} (\bibinfo {year} {2003}{\natexlab{a}})},\ \Eprint
  {http://arxiv.org/abs/hep-ph/0301185} {arXiv:hep-ph/0301185 [hep-ph]}
  \BibitemShut {NoStop}%
\bibitem [{\citenamefont {Grigoriev}\ \emph
  {et~al.}(2003{\natexlab{b}})\citenamefont {Grigoriev}, \citenamefont
  {Jankowski},\ and\ \citenamefont {Tkachov}}]{Grigoriev:2003tn}%
  \BibitemOpen
  \bibfield  {author} {\bibinfo {author} {\bibfnamefont {D.}~\bibnamefont
  {Grigoriev}}, \bibinfo {author} {\bibfnamefont {E.}~\bibnamefont
  {Jankowski}}, \ and\ \bibinfo {author} {\bibfnamefont {F.}~\bibnamefont
  {Tkachov}},\ }\href@noop {} {\bibfield  {journal} {\bibinfo  {journal}
  {Comput.Phys.Commun.}\ }\textbf {\bibinfo {volume} {155}},\ \bibinfo {pages}
  {42} (\bibinfo {year} {2003}{\natexlab{b}})},\ \Eprint
  {http://arxiv.org/abs/hep-ph/0301226} {arXiv:hep-ph/0301226 [hep-ph]}
  \BibitemShut {NoStop}%
\bibitem [{\citenamefont {Chekanov}(2006)}]{Chekanov:2005cq}%
  \BibitemOpen
  \bibfield  {author} {\bibinfo {author} {\bibfnamefont {S.}~\bibnamefont
  {Chekanov}},\ }\href@noop {} {\bibfield  {journal} {\bibinfo  {journal}
  {Eur.Phys.J.}\ }\textbf {\bibinfo {volume} {C47}},\ \bibinfo {pages} {611}
  (\bibinfo {year} {2006})},\ \Eprint {http://arxiv.org/abs/hep-ph/0512027}
  {arXiv:hep-ph/0512027 [hep-ph]} \BibitemShut {NoStop}%
\bibitem [{\citenamefont {Lai}\ and\ \citenamefont {Cole}(2008)}]{Lai:2008zp}%
  \BibitemOpen
  \bibfield  {author} {\bibinfo {author} {\bibfnamefont {Y.-S.}\ \bibnamefont
  {Lai}}\ and\ \bibinfo {author} {\bibfnamefont {B.~A.}\ \bibnamefont {Cole}},\
  }\href@noop {} {\  (\bibinfo {year} {2008})},\ \Eprint
  {http://arxiv.org/abs/0806.1499} {arXiv:0806.1499 [nucl-ex]} \BibitemShut
  {NoStop}%
\bibitem [{\citenamefont {Volobouev}(2009)}]{Volobouev:2009rv}%
  \BibitemOpen
  \bibfield  {author} {\bibinfo {author} {\bibfnamefont {I.}~\bibnamefont
  {Volobouev}},\ }\href@noop {} {\  (\bibinfo {year} {2009})},\ \Eprint
  {http://arxiv.org/abs/0907.0270} {arXiv:0907.0270 [hep-ex]} \BibitemShut
  {NoStop}%
\end{thebibliography}%

\end{document}